# Evaluating the severity of tandem coronary stenoses: Insights from simulated FFR and iFR techniques

*Navid Freidoonimehr[1], Tam Atkins[1], Jessica A. Marathe[2,3,4], Peter J. Psaltis[2,3,4], and Maziar Arjomandi[1]*

[1] *School of Electrical and Mechanical Engineering, The University of Adelaide, South Australia 5005 Australia*

[2] *Vascular Research Centre, Lifelong Health Theme, South Australian Health and Medical Research Institute (SAHMRI), Adelaide, South Australia 5000, Australia*

[3] *Adelaide Medical School, University of Adelaide, Adelaide, South Australia 5005, Australia*

[4] *Department of Cardiology, Central Adelaide Local Health Network, Adelaide, South Australia 5000, Australia*

---

[1] Email: navid.freidoonimehr@adelaide.edu.au

# Abstract

**Aims:** The presence of coronary tandem lesions poses a significant challenge for the accurate diagnosis and management of coronary artery diseases. This study set out to provide a deeper understanding of the haemodynamic interactions between tandem obstructive coronary lesions and their impact on different haemodynamic diagnostic parameters.

**Methods and Results:** Using a computational fluid dynamic model, validated against *in vitro* laboratory experiments, we investigated the various combinations of moderate and severe stenoses interchangeably located in the proximal and distal segments of the artery. The investigation was conducted using two diagnostic parameters: one hyperaemic-based, i.e., FFR, and one rest-based, i.e., iFR, technique, both of which are commonly used to assess the physiological significance of coronary stenoses. The three main findings of this work are: (a) the recovery distance (the immediate local distance affected by the presence of stenosis) is much shorter for the rest-measured diagnostic parameter compared with the hyperaemic-measured diagnostic parameter; (b) pressure drop measurements immediately downstream of the stenotic sections overestimate the significance of stenoses, and (c) the presence of a moderate stenosis downstream of a severe stenosis increases FFR value (faster FFR recovery).

**Conclusion:** These findings enhance our understanding of the diagnostic accuracy of hyperaemic-based and rest-based physiological diagnostic coronary assessments and the nuances of using these different techniques when assessing tandem coronary stenoses. This understanding can help inform tailored therapeutic approaches for the management of coronary artery disease.

# 1 Introduction

The presence of tandem stenoses—two or more sequential narrowings — within a coronary artery presents a unique challenge for understanding and managing coronary artery diseases (CAD)[1]. These complex lesions can complicate the interpretation of coronary physiology, making it difficult to determine the optimal treatment strategy. Traditional physiological assessment techniques, such as Fractional Flow Reserve (FFR)[2, 3] and Instantaneous Wave-Free Ratio (iFR)[4], have proven valuable in evaluating single stenoses[5-8]; however, their accuracy can be compromised when applied to tandem stenoses. In current clinical practice, FFR and iFR are employed to assess the haemodynamic significance of lesions, yet their limitations in the context of tandem stenoses can lead to suboptimal treatment decisions. These limitations include but are not limited to complex decision-making on which lesions to treat and in what order as well as serial lesions interaction which influences coronary haemodynamics in a nonlinear manner.

The prevalence of tandem coronary stenoses has been reported to be between 20-40% in patients with CAD presenting for invasive coronary angiography[9, 10]. Obstructive tandem stenoses (with degrees of stenosis of more than 50%[11]) introduce additional complexity to the physiological assessment of coronary disease, due to the interaction between proximal and distal lesions, which can significantly alter flow patterns and pressure gradients[12, 13]. This can make it difficult to accurately identify the physiological significance of individual lesions using the aforementioned physiologically-based diagnostic techniques[14]. As noted by Modi et al.[15], the application of FFR to assess the functional significance of individual stenoses in serially diseased vessels is theoretically limited because of the presence of additional resistance to flow downstream, with the situation further complicated by flow turbulence when stenoses are particularly severe, close together and non-concentric. In addition, it was shown that the individual stenoses were considerably underestimated in the serial arrangements[16]. To overcome these challenges, attempts have been made in the literature by introducing mathematical correction models to assess the individual significance of each stenosis when tandem lesions are present[16-20].

To address these shortcomings and provide a deeper insight into the haemodynamic assessment of tandem coronary stenoses, we have undertaken a study combining computational fluid dynamic modelling (CFD) and in vitro experiments. Understanding the pressure drop across tandem stenoses is crucial for several reasons: first, it provides insight into the physiological impact of these lesions on coronary blood flow and myocardial oxygen supply in the vessel territory. Second, it aids in the development of more accurate diagnostic tools and criteria for assessing the severity of CAD, particularly in complex cases with multiple lesions. Finally, insights into the haemodynamic effects of tandem stenoses can guide therapeutic decision-making such as the decision to revascularise both lesions with stenting or only one lesion. We therefore set out to investigate the effect of tandem stenoses on pressure drop within the coronary arteries by simulating and comparing two invasive physiologically-based diagnostic techniques - FFR and iFR - to investigate the differences in their ability to interpret the significance of such lesions.

## 2 Methods

### 2.1 Coronary artery model with tandem stenoses

The coronary artery models developed and used in this study were based on the examples of tandem stenoses seen on invasive coronary angiography. An idealised coronary artery model with a circular cross-section was adapted from this (Figure 1) to allow parameters of interest to be varied, in lieu of using fixed patient-specific geometry. In addition, although the coronary arteries are naturally tapered[21], we modelled the artery as a straight cylinder for this study. This was to allow us to focus solely on the impact of tandem stenoses on the haemodynamics. The artery diameter, $D$, was set at 3 mm, and the total artery length was fixed at 150 mm ($50D$). The length of each stenosis was fixed at $2D$[22], and the distance between the two stenoses was denoted as $L$. Both stenoses were asymmetric with a circular cross-section at various degrees of stenosis, defined as the ratio of narrowed to normal segments.

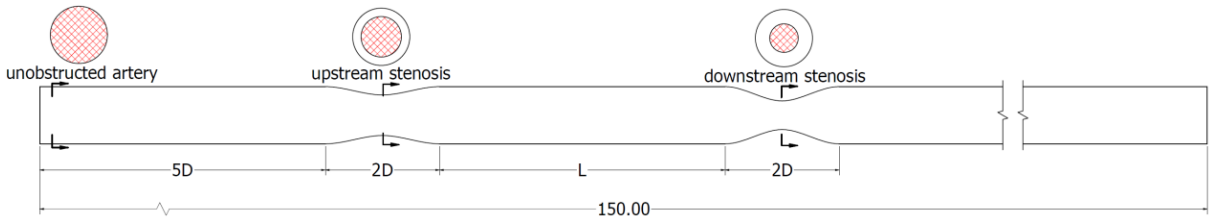

Figure 1: Schematic of the simplified coronary artery model with tandem stenoses. The artery diameter and the total length of the artery are fixed at 3 mm and 150 mm (50D), respectively, for all simulations.

### 2.2 Computational modelling

ANSYS FLUENT 2023 R1 (ANSYS, Inc) was employed to simulate the blood flow within the coronary artery model featuring tandem stenoses. The solver settings to solve the governing incompressible Navier–Stokes equations have been previously described[23]. Parabolic velocity profiles with the average flow rate of 1.1 ml/s and 3.9 ml/s, corresponding to the mean wave-free period at rest and the mean complete cardiac cycle at hyperaemia flow rates, respectively, were used as the inflow boundary condition to represent fully developed blood flow (Figure 2). A time step of $5\times10^{-6}$ was selected to ensure the Courant number was much less than unity. Furthermore, a pressure outlet boundary condition with a physiological pressure value of 80 mmHg and a no-slip boundary condition at the walls were considered. The blood was modelled as a Newtonian fluid, a valid assumption in large arteries[24], with a density of 1060 kg/m³ and a dynamic viscosity of 3.4 mPa.s.

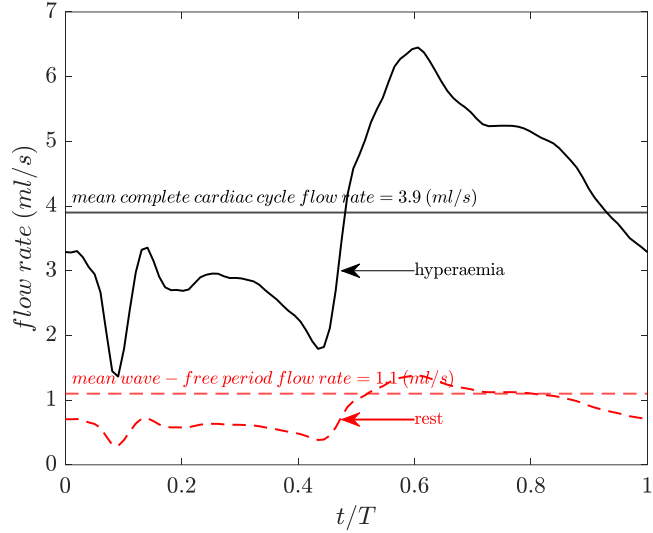

*Figure 2: Physiological flow waveforms of a coronary artery. The profiles are shown at rest and hyperaemia as well as their equivalent mean complete cardiac cycle flow rate at hyperaemia (horizontal solid black line) and the mean wave-free period flow rate at rest (horizontal red-dashed line).*

We employed ANSYS ICEM CFD for meshing the model using block-structured elements (Figure 3a). The choice of structured elements is due to their faster convergence compared to unstructured mesh elements. To ensure the independence of the pressure drop and FFR results from the number of elements, we conducted a convergence study on the required number of mesh elements for each stream-wise, cross-sectional, and near-wall direction (Figure 3b-d). In each direction, the criteria to choose the optimum element size were set such that the variations in pressure drop values would be less than 0.5%. Hence, the optimum element size in the stream-wise direction was selected as $2\times10^{-4}$, in the cross-sectional direction as $1.5\times10^{-4}$, with 20 layers of mesh elements near the arterial wall, and with a total of about 1.34 million elements.

## 2.3 CFD validation

To validate the results obtained from CFD simulations, we conducted a series of pressure measurement experiments using the coronary arterial experimental apparatus details of which have been previously described by the authors[25]. A programmable magnet drive gear pump (Fluid-o-Tech, FG404) was used to create steady and pulsatile flows. An electromagnetic flow meter (Krohne BATCHFLUX 5500 C) was mounted to record the flow rate. The apparatus was equipped with three pressure transducers (Druck UNIK 5000) to measure the pressure upstream of the first stenosis, between the two stenoses, and downstream of the second stenosis. While ensuring the consistency of the Reynolds and Womersley numbers, which are two dimensionless numbers governing physiological flows, between the experiments and CFD simulations, the coronary artery models were 3D-printed with an internal diameter of 6 mm. This diameter is double that of the average coronary artery and the one developed in the CFD simulations. This adjustment was made for compatibility with the rest of the experimental apparatus and to ease the selection of suitable sensors and data collection. To compare the results obtained from CFD and experiments, the pressure drop across each stenosis and across both stenoses in a tandem arrangement was non-dimensionalised using the dynamic pressure calculated based on the bulk flow velocity. Consequently, the pressure drop coefficient ($C_{\Delta P}$) was introduced as the pressure drop, $\Delta P$, divided by the dynamic pressure ($\frac{1}{2}\rho U^2$ where $\rho$ is the fluid density and $U$ is the un-obstructed bulk flow velocity) - i.e. $C_{\Delta P_{first}} = \frac{P_1 - P_2}{\frac{1}{2}\rho U^2}$.

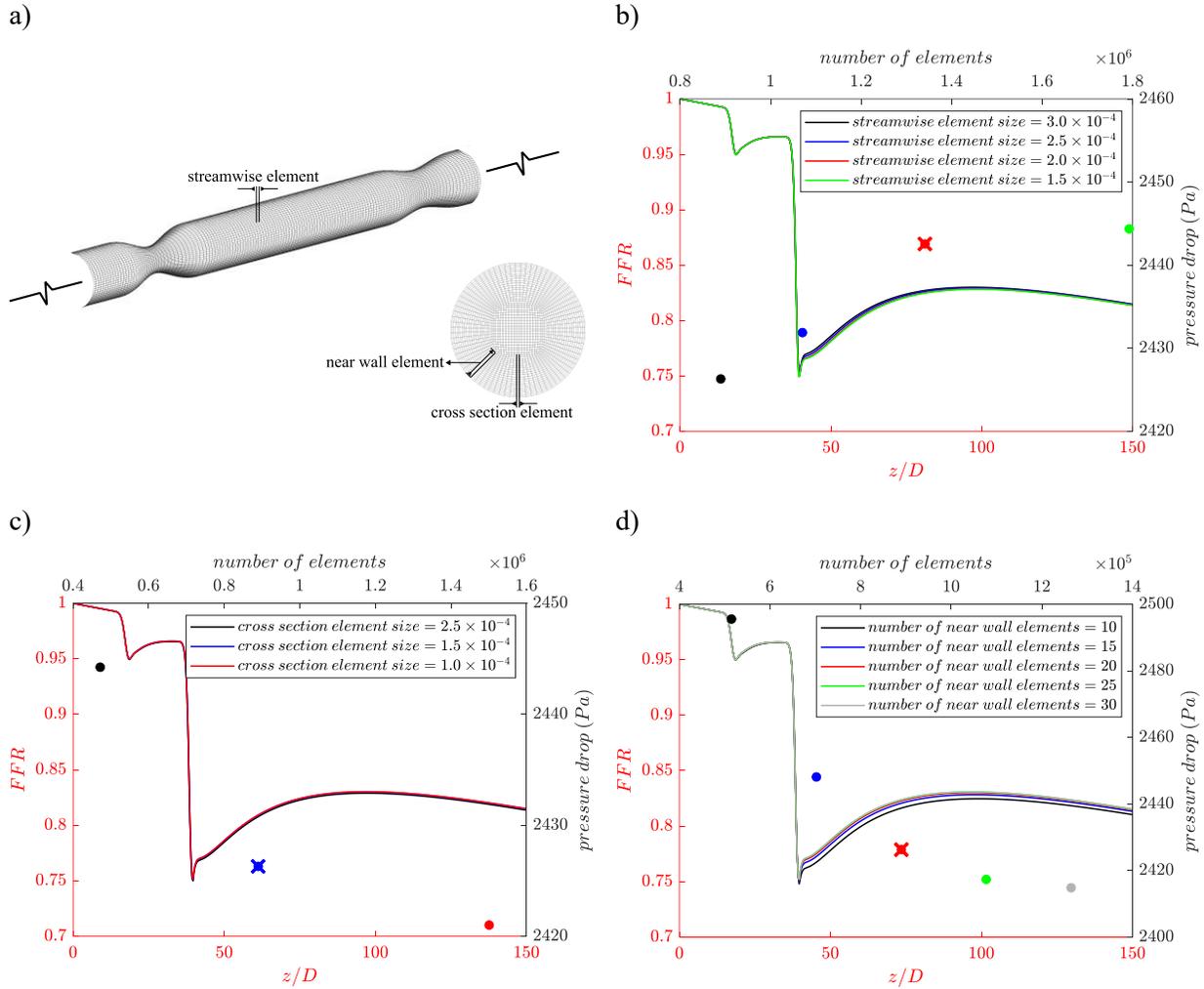

Figure 3: Mesh design used for CFD modelling. Figure 3(a) shows the schematic of the mesh structure highlighting stream-wise, cross-sectional, and near-wall elements. Variations of FFR (profiles with respect to the left vertical axis) and pressure drop (symbols with respect to the right vertical axis) along the stream-wise direction are shown for different stream-wise element sizes in figure 3(b), cross-sectional element sizes in figure 3(c), and near-wall elements in figure 3(d). The 'cross' symbols indicate the chosen element sizes in different directions for which the variations in pressure drop are less than 0.5%.

# 3 Results

## 3.1 Accuracy of CFD simulations

To validate the CFD modelling, we compared the pressure drop coefficients obtained from CFD with laboratory experiments for three stream-wise locations (Figure 4). The experiments were conducted for both steady and pulsatile flows with the same average flow velocity. The results from the pulsatile experiments were used to validate the steady CFD simulation results by calculating the percentage changes in the pressure drop coefficient (Table 1). Our comparison showed that the CFD results closely match the experimental data, with the average changes in pressure drop coefficients obtained from steady CFD and pulsatile experiments being 11%.

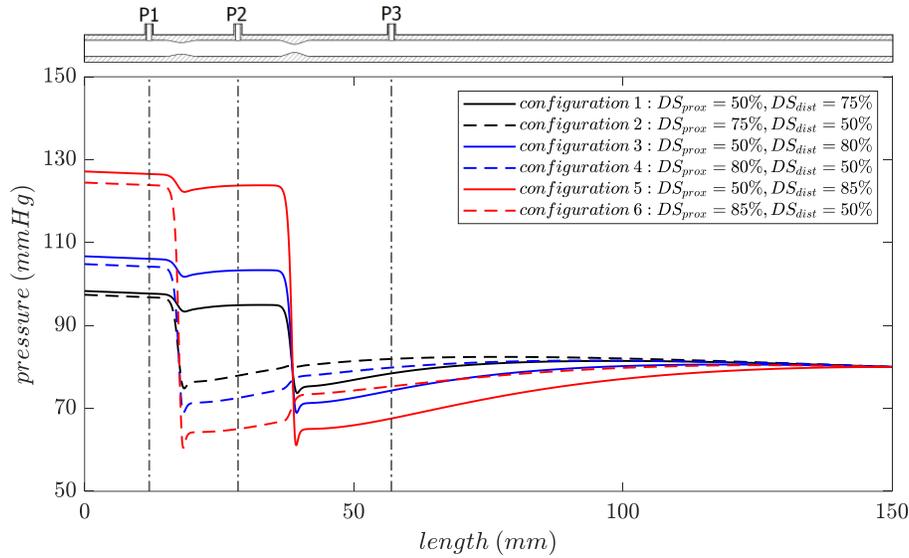

*Figure 4: Pressure profiles obtained from CFD used for validation against laboratory experiments. Figure highlights the locations of experimental pressure ports, P1, P2, and P3 (vertical dash-dotted lines). The distance between the two stenoses is fixed at 5D (15 mm). The first pressure port, P1, is located at 1D (3 mm) before the first stenosis; the second port, P2, is located in the middle of the two stenoses; and the third port, P3, is located at 5D after the second stenosis.*

Table 1: Comparison between CFD modelling and laboratory experiment results. The pressure drop coefficients are tabulated across the first, second, and both stenoses during hyperaemia obtained from CFD and pressure measurement experiments, both for steady ('stead') and pulsatile ('puls') flows, for different configurations, as specified in Figure 4. The '_' signs indicate that the experimental data is not available.

| $C_{\Delta P}$ | configuration 1 | | | | configuration 2 | | | | configuration 3 | | | | configuration 4 | | | | configuration 5 | | | | configuration 6 | | | |
|---|---|---|---|---|---|---|---|---|---|---|---|---|---|---|---|---|---|---|---|---|---|---|---|---|
| | CFD | exp | | change | CFD | exp | | change | CFD | exp | | Change | CFD | exp | | change | CFD | exp | | change | CFD | exp | | change |
| | | stead | puls | | | stead | puls | | | stead | puls | | | stead | puls | | | stead | puls | | | stead | puls | |
| $C_{\Delta P_{first}}$ | 2.3 | 2.4 | _ | _ | 15.6 | 17.5 | 18.6 | **15.8%** | 2.3 | _ | 2.5 | **7.8%** | 26.2 | 22.4 | 26.1 | **0.2%** | 2.3 | _ | _ | _ | 48.6 | 48.7 | 51.6 | **5.8%** |
| $C_{\Delta P_{second}}$ | 13.6 | 11.7 | 16.3 | **8.2%** | -3.3 | -3.1 | -2.7 | **24.2%** | 23.9 | 23.2 | 22.2 | **8%** | -6.1 | _ | _ | _ | 46.4 | 39.6 | 43.9 | **5.8%** | -8.5 | _ | _ | _ |
| $C_{\Delta P_{both}}$ | 15.9 | 14 | 13.3 | **13%** | 12.3 | 14.4 | 15.9 | **22.6%** | 26.3 | 24 | 24.7 | **6.4%** | 20.1 | 19.4 | 23 | **12.4%** | 48.8 | 44.6 | 45.8 | **6.5%** | 40.1 | 44.3 | 47.4 | **15.3%** |

## 3.2 Effect of intra-distance between two stenoses

We first investigated the effect of the distance between two stenoses, or intra-distance, on the distal to proximal pressure ratio at rest (representative of the wave-free ratio) and FFR (Figure 5). Both profiles featured a region immediately downstream of the stenotic sections which is locally influenced by the presence of stenoses. This region is called the recovery zone and was characterised by high-velocity areas immediately downstream of the stenosis, called jet regions, surrounded by recirculation zones where the flow is separated (Figure 6). We observed that the length of these zones is a function of the severity of the most dominant stenosis and the presence of hyperaemia or rest events.

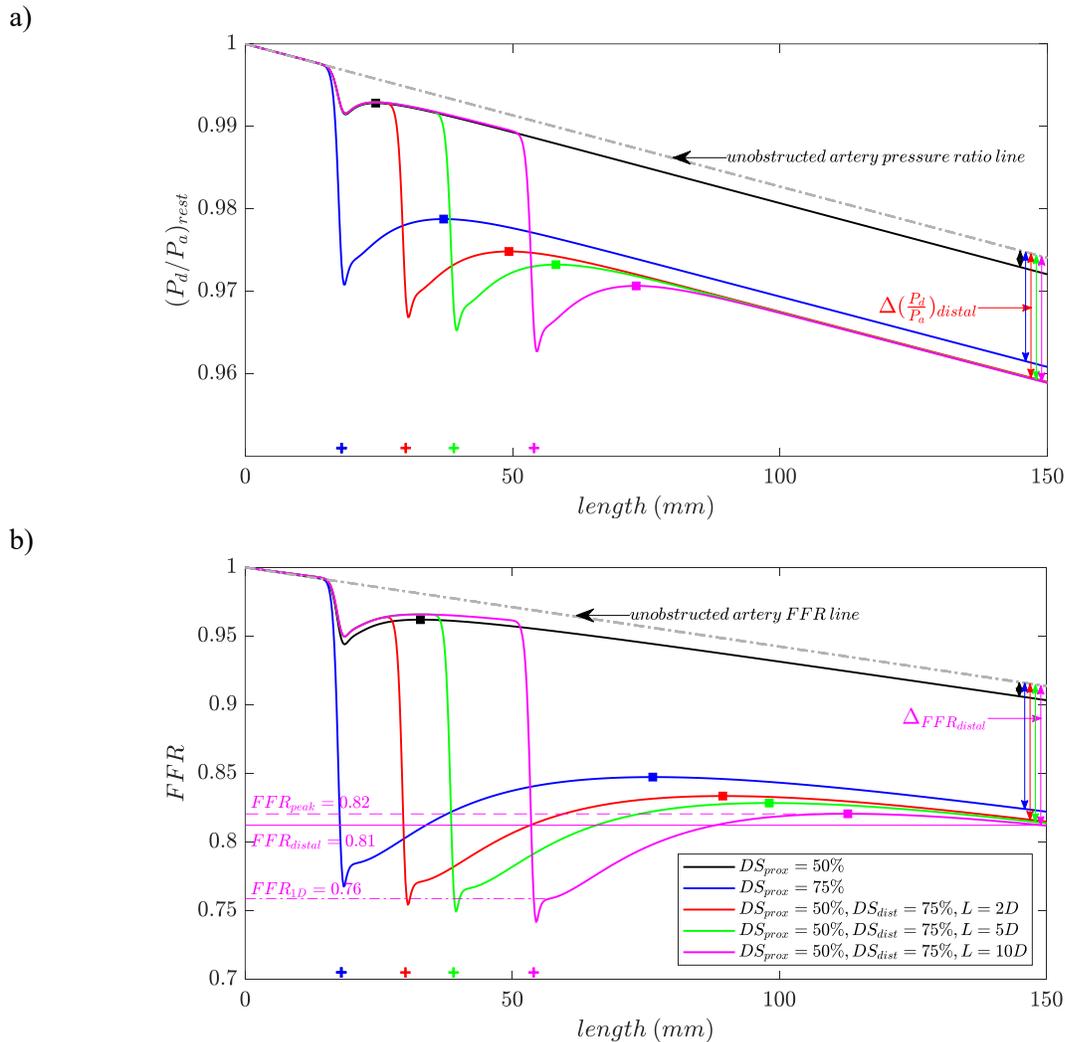

Figure 5: Effect of the distance between two tandem stenoses. Figure 5(a) shows the profile of distal to proximal pressure ratio at rest and figure 5(b) shows the FFR profiles. The 'square' symbols indicate the locations after the stenoses where the FFR profiles are almost no longer influenced by the presence of stenosis. The dotted-dashed grey line shows the FFR profile of an unobstructed artery model during hyperaemia. The 'plus' signs on the bottom horizontal axis indicate the locations of the stenoses.

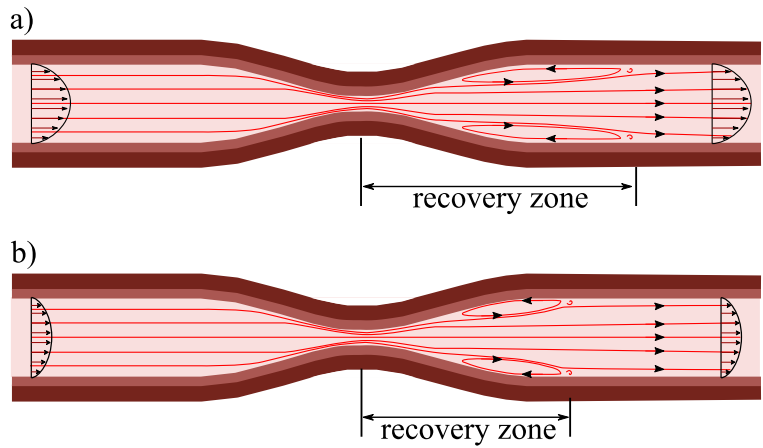

*Figure 6: A schematic of the flow downstream of the stenotic section highlighting flow recirculation zones, turbulence eddies, and recovery zones at (a) hyperaemia and (b) rest.*

### 3.3  Effect of degree of stenosis on the recovery distance

We then examined the effect of the degree of stenosis on FFR and rest-measured pressure ratio, as well as the recovery distance required after the stenosis (Figure 7). This was achieved based on the analysis of single stenosis, as we showed that the diagnostic parameters are mainly dominated by the effect of more severe stenosis (Figure 5). We found that the recovery length - the required distance for diagnostic parameters to resume their decreasing behaviour after stenosis throughout the artery - is directly proportional to the pressure drop (Figures 7(c) and (d)).

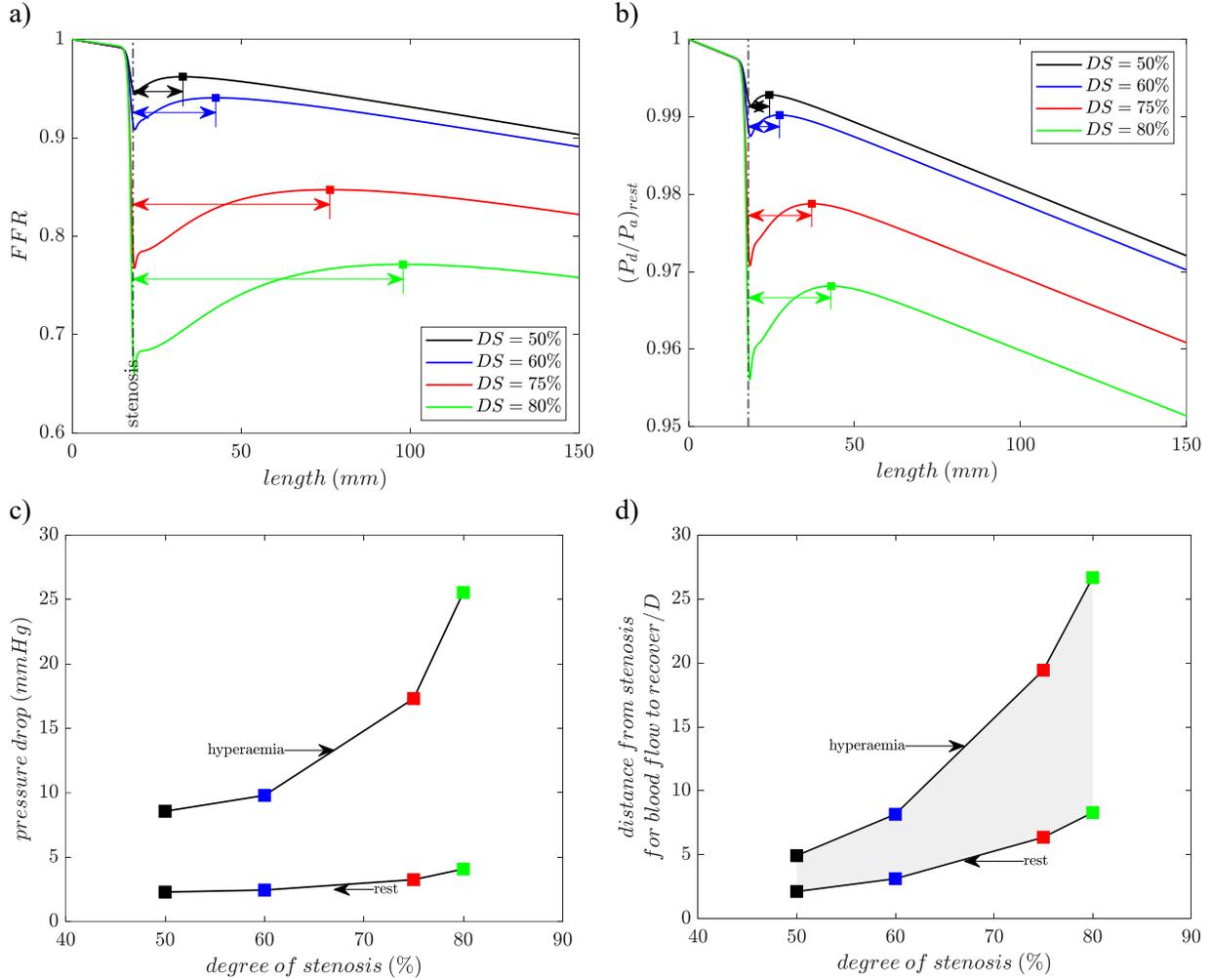

*Figure 7: Effect of the degree of stenosis (DS) on the recovery distance. The effect of DS on FFR (figure 7(a)) and distal to proximal pressure ratio measured at rest (figure 7(b)) throughout the artery are presented highlighting the recovery distance required after the stenosis (horizontal double-ended arrow line) for accurate FFR and pressure ratio measurement. Figure 7(c) shows the pressure drop and figure 7(d) shows the normalised recovery distance from the location of stenosis versus the degree of stenosis at rest and hyperaemia. The grey-shaded area in figure 7(d) shows the region of recovery distance as a function of the degree of stenosis and conditions from rest to hyperaemia.*

### 3.4  Effect of varying the severity of the more severe stenosis

We also investigated the effect of variations in the severity of the more severe stenosis in a tandem arrangement and their stream-wise locations on FFR (Figure 8). We observed that when the degree of stenosis of the more severe stenosis located downstream increases, the FFR profiles experience a more intense drop at the location of the severe stenosis (Figure 8(a)). We then investigated the FFR behaviour when the more severe stenosis with various degrees of stenosis is located upstream, followed by a constant moderate stenosis (50%) (Figure 8(b)). A local increase in the FFR profiles was observed at the location of the moderate downstream stenosis, particularly in the presence of severe stenosis upstream.

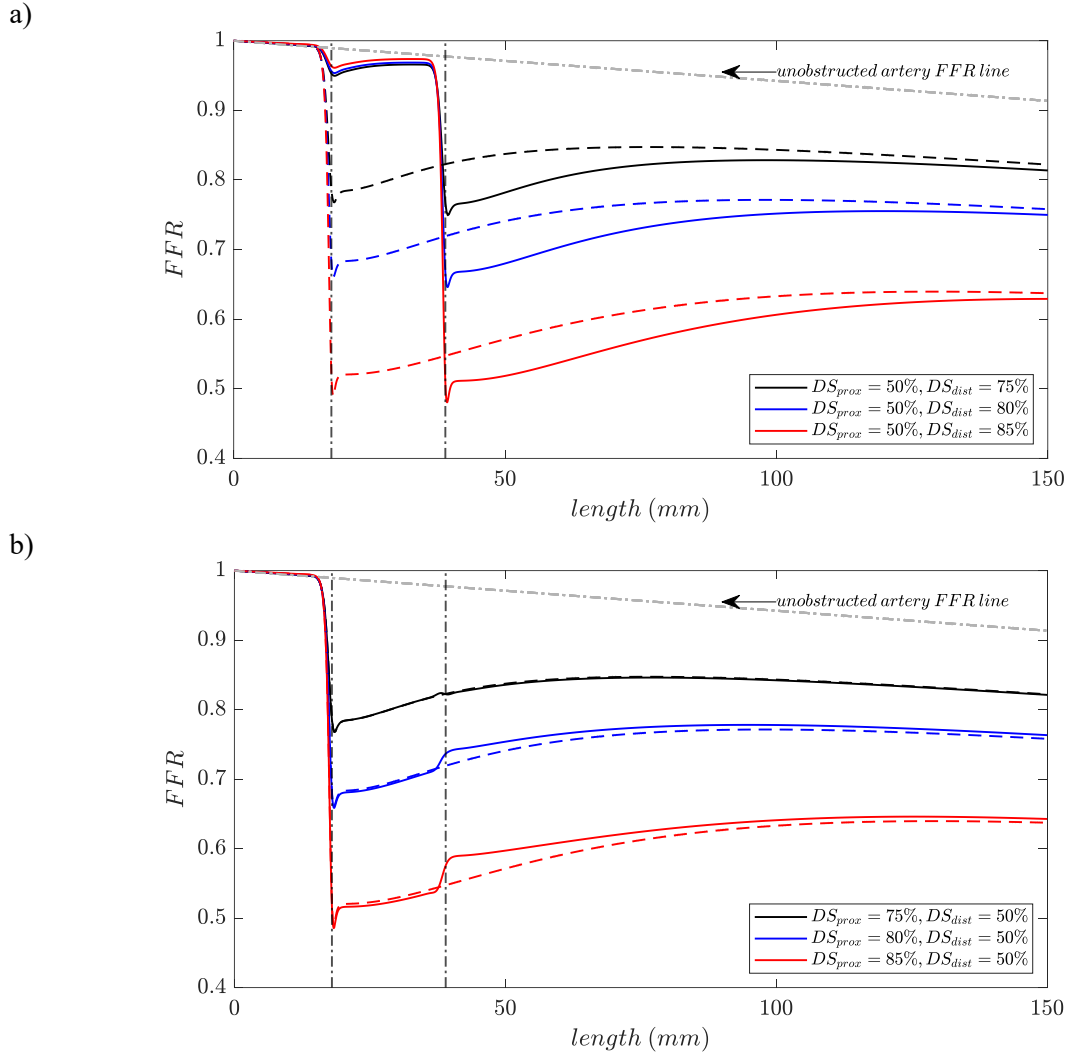

*Figure 8: Effect of varying the severity of the more severe stenosis. FFR profiles throughout the artery with tandem stenoses, with severe stenosis located (a) downstream and (b) upstream. The vertical dashed-dotted lines indicate the location of the stenoses. The FFR profile of an unobstructed artery is plotted for reference. The dashed lines refer to the FFR profiles for the artery model with a single stenosis.*

## 4  Discussion

This study has identified three main findings relating to the impact of tandem lesions on coronary haemodynamics. The first was the introduction of the recovery zone and its role in influencing the physiological significance of each stenosis. The recovery zone was introduced to refer to the area immediately downstream of a stenosis where the flow is disturbed due to the formation of flow recirculation zones and flow turbulences. As the same unobstructed upstream flow needs to pass through the obstructed area, a core region with high velocity as a part of the recovery zone is created. It is known that the total pressure and kinetic energies of the fluid must remain constant along a flow streamline (Bernoulli's principle[26]). Based on this principle, for blood flow in an artery, because the area immediately downstream of stenosis encompasses a zone with high velocity and consequently higher kinetic energy, the pressure must be considerably lower in this zone. Therefore, measurement of pressure (or other pressure-based diagnostic parameters, e.g. FFR or iFR) in this zone might result in overestimation of the significance of a given stenosis. For example, we determined that the FFR measurement of tandem stenoses with 50% upstream and 75% downstream stenoses with a $10D$ distance between them results in a 6% difference in the values measured immediately downstream compared to the farthest distance possible. A local maximum in the rest pressure ratio and FFR profiles downstream of the stenoses can be observed, representing the location where the jet flow region and recirculation zones combine, representing the recovery zone settlement. Beyond this local maximum, the profiles tend to achieve the same slope as the unobstructed artery profiles. A similar trend was reported by Li et al.[27] for an angiography-derived FFR of a coronary artery with an intermediate stenosis. Beyond this point, the FFR values become close to those measured at the farthest downstream of the artery. For example, only 1% deviations can be observed for FFR measurement at local maxima (~0.82) compared to the farthest distance (~0.81). Hence, our findings suggest that for accurate determination of the significance of each stenosis, the measurements should be conducted at the furthest possible distance downstream of the stenosis, with the optimal threshold distance a function of the degree of stenosis. In addition, we show that the location of local maxima shifts further downstream with increasing intra-distance with the extent of the shift directly proportional to the variations in the distance between the two stenoses. However, the profiles of different intra-distances collapse at the farthest downstream if enough downstream length is available. Therefore, the variation in intra-distance does not change the measured rest pressure drop and FFR values very distally in the artery.

Our second main finding relates to the comparison between hyperaemic- and rest-measured diagnostic parameters. Upon administration of vasodilators to induce hyperaemia, the flow rate within the coronaries increases and hence the flow recirculation zones and area with high velocity become stronger/larger downstream of the stenotic sections compared to under the rest condition. Hence, the recovery zone becomes longer in hyperaemia. For example, the diameter-normalised recovery distance for a configuration with a degree of stenosis of 60% at rest is about 3, while at hyperaemia it is about 8. This indicates that for the physiological assessment of coronary stenosis severity, hyperaemic measurements need to be performed further downstream of the stenosis than under rest conditions. This implies that if the stenosis is located towards the end of the artery of interest and/or if the downstream distance beyond the stenosis is limited, rest-measured parameters (e.g., with iFR) could provide a more accurate representation of the functional significance of stenoses compared to hyperaemic-based ones.

The third new insight from this study regards the impact of the arrangement of moderate and severe stenoses in tandem configuration. The presence of a moderate downstream stenosis, when paired with a severe upstream stenosis, could lead to a faster recovery of FFR profiles and thus enhance blood circulation within the artery. This suggests that in the discussed arrangement, the presence of a downstream moderate stenosis could offer more benefits than disadvantages from a physiological perspective. For example, in the tandem arrangement with an upstream degree of stenosis of 85% and downstream of 50%, the FFR profile rises from about 0.54 to 0.59 at the location of the moderate downstream stenosis. This suggests that for more severe upstream stenosis (i.e., 80% and 85%), the downstream moderate stenosis can locally re-energise the flow. This assists the FFR profiles in recovering faster to their corresponding single stenosis FFR profile. However, from another perspective, a recent study by Franke et al.[28] provides insight into the potential clinical implications of having tandem coronary stenoses. The authors showed that coronary lesions that have a downstream stenosis in the same artery could be at an increased risk of stenosis progression. These findings together with our current results concerning the recovery zone, highlight the potential importance of coronary biomechanics on atherosclerotic disease progression[29], and further work on the practical implications of having tandem coronary stenoses is required.

Limitations of our study include the use of a simplified model of a coronary artery with tandem stenoses instead of patient-specific geometries. However, as the objective of this work was to investigate different configurations of tandem lesions, the choice of a simplified artery model was essential. The simplifications include not considering the effects of arterial tapering, branches, and microvasculature. The first two simplifications were to isolate the effect of stenoses and their varying severities on the haemodynamics and the latter is unlikely to have affected our results as the effect of increased microvascular resistance should be consistent throughout the length of the artery.

In conclusion, we investigated the impact of stenosis severity on the FFR and rest-measured pressure ratio measurements in arterial models with tandem stenoses. Our findings emphasise the significant influence of the severity of each stenosis and its position on diagnostic parameters and specifically highlight the dominant role that the most severe stenosis plays in modifying haemodynamic parameters, irrespective of its position in tandem arrangements. In addition, we showed that measurements taken immediately downstream of stenoses could differ significantly from those taken further downstream, with a notable recovery zone where FFR and rest pressure ratios begin to stabilise. We depicted that this recovery zone for diagnostic parameters measured at rest (e.g. iFR) was much shorter compared to those measured during hyperaemia (FFR), at a similar degree of stenosis, implying that the iFR measurement could provide a more accurate representation of the functional significance of stenoses compared to FFR, specifically where the distal distance is limited. These insights could enhance diagnostic accuracy and inform more tailored therapeutic approaches, as well as highlight the need for further clinical and fluid mechanics studies to explore the full potential of these findings in the context of cardiovascular health management.

## Funding

This work was supported by Level 3 Future Leader Fellowships (P.J.P) from the National Heart Foundation of Australia [FLF 106656].

## Acknowledgements

This work was supported with supercomputing resources provided by the Phoenix HPC service at the University of Adelaide.

## Conflict of Interest:

P.J.P. reports research support from Abbott Vascular, consulting fees from Amgen, Eli Lilly, Novo Nordisk, Esperion and Novartis, and speaker honoraria from Amgen, AstraZeneca, Bayer, Boehringer Ingelheim, Merck Schering-Plough, Pfizer, Novartis, Novo Nordisk and Sanofi.